\documentclass[aps,prl,notitlepage,superscriptaddress,reprint,nofootinbib]{revtex4-1}
\usepackage{graphics}
\usepackage[tbtags]{amsmath}
\usepackage{amssymb,amsthm} 
\usepackage{epstopdf,cancel,ulem}
\usepackage{epsf,latexsym,bbm,euscript}
\usepackage[colorlinks,bookmarks=false]{hyperref}
\newcommand {\h}[1]{\hat{ #1}}
\def\hint{\hat H_{int}}

\newcommand{\cH}{{\cal H}}

\def\go{g_{00}}

\newcommand{\be}{\begin{equation}}
\newcommand{\ee}[1]{\label{#1} \end{equation}}
\newcommand{\bbe}{\begin{equation*}}
\newcommand{\eee}{\end{equation*}}

\newcommand{\bs}{\begin{split}}
\newcommand{\es}{\end{split}}

\newcommand{\e}[1]{e^{{ \textstyle #1 }}}

\newcommand {\ket}[1]{\lvert \, #1\rangle}
\newcommand {\bra}[1]{\langle #1 \, \rvert}
\newcommand {\braket}[2]{\langle #1 \, | \, #2 \rangle}
\newcommand {\proj}[1]{| #1\rangle \langle #1 |}

\def\Tr{ {\textrm{Tr} }}

\def\f12{\frac{1}{2}}

\newcommand{\mean}[1]{\ensuremath{\left\langle #1 \right\rangle}}

\usepackage{comment} 
\usepackage{color} 
\usepackage{cancel}

\usepackage[utf8]{inputenc}
\usepackage{siunitx} 
\usepackage{braket}
\usepackage[tiny]{titlesec}

\titleformat*{\section}{\large\bfseries}
\titleformat*{\subsection}{\normalsize \bfseries}
\titleformat*{\subsubsection}{\normalsize\bfseries}
\usepackage{xspace}
\usepackage{times}
\usepackage{bm}			
\usepackage{amsfonts}
\usepackage{xspace}
\usepackage{epsfig}
\usepackage{sidecap}
\usepackage[titletoc,toc,title]{appendix}
\usepackage{dsfont} 

\begin{document}

\newcommand{\eq}[1]{Eq.\,(\ref{#1})\xspace}
\newcommand{\Shat}{\ensuremath{\hat{S}}\xspace}
\newcommand{\vj}{\ensuremath{|+ -\rangle}\xspace}
\newcommand{\bj}{\ensuremath{|- +\rangle}\xspace}
\newcommand{\hbs}{\ensuremath{\frac{\hbar^2}{4}}\xspace}
\newcommand{\thbs}{\ensuremath{\frac{2\hbar^2}{4}}\xspace}
\newcommand{\hbss}{\ensuremath{\frac{3\hbar^2}{4}}\xspace}
\newcommand{\pp}{\ensuremath{|\psi_2\rangle}\xspace}
\newcommand{\adag}{\ensuremath{\hat{a}^{\dagger}}\xspace}
\newcommand{\jx}{\ensuremath{\hat{J}_x}\xspace}
\newcommand{\jy}{\ensuremath{\hat{J}_y}\xspace}
\newcommand{\jz}{\ensuremath{\hat{J}_z}\xspace}
\newcommand{\jk}{\ensuremath{\hat{J}_+}\xspace}
\newcommand{\jl}{\ensuremath{\hat{J}_-}\xspace}
\newcommand{\ef}{\ensuremath{\epsilon_0}\xspace}
\newcommand{\gam}{\ensuremath{\frac{1}{\sqrt(1-\beta^2)}}\xspace}
\newcommand{\invgam}{\ensuremath{\sqrt{1-\beta^2}}\xspace}
\newcommand{\muf}{\ensuremath{\mu_0}\xspace}
\newcommand{\boldE}{\ensuremath{\mathbf{E}}\xspace}
\newcommand{\boldB}{\ensuremath{\mathbf{B}}\xspace}
\newcommand{\boldJ}{\ensuremath{\mathbf{J}}\xspace}
\newcommand{\boldD}{\ensuremath{\mathbf{D}}\xspace}
\newcommand{\boldH}{\ensuremath{\mathbf{H}}\xspace}
\newcommand{\boldj}{\ensuremath{\mathbf{j}}\xspace}
\newcommand{\Fr}{\ensuremath{\mathbf{F}_{rad}}\xspace}
\newcommand{\bH}{\ensuremath{\mathbf{H}}\xspace}
\newcommand{\boldF}{\ensuremath{\mathbf{F}}\xspace}
\newcommand{\E}{\ensuremath{\mathbf{E}}\xspace}
\newcommand{\bv}{\ensuremath{\mathbf{v}}\xspace}
\newcommand{\fu}{\ensuremath{\mathbf{F}^{\mu}}\xspace}
\newcommand{\ful}{\ensuremath{\mathbf{F}_{\mu}}\xspace}
\newcommand{\fm}{\ensuremath{\mathbf{F}^{m}}\xspace}
\newcommand{\fobs}{\ensuremath{\mathbf{F}^{obs}}\xspace}
\newcommand{\fobsl}{\ensuremath{\mathbf{F}_{obs}}\xspace}
\newcommand{\wobs}{\ensuremath{\omega_{obs}}\xspace}
\newcommand{\wm}{\ensuremath{\omega_{m}}\xspace}
\newcommand{\kxobs}{\ensuremath{k_x}\xspace}
\newcommand{\kyobs}{\ensuremath{k_{y}}\xspace}
\newcommand{\kzobs}{\ensuremath{k_{z}}\xspace}
\newcommand{\kxm}{\ensuremath{\bar{k}_x}\xspace}
\newcommand{\kym}{\ensuremath{\bar{k}_{y}}\xspace}
\newcommand{\kzm}{\ensuremath{\bar{k}_{z}}\xspace}
\newcommand{\bg}{\ensuremath{-\beta\gamma}\xspace}
\newcommand{\gu}{\ensuremath{\mathbf{G}^{\mu}}\xspace}
\newcommand{\vs}[1]{\ensuremath{\boldsymbol{#1}}\xspace}
\renewcommand{\v}[1]{\ensuremath{\mathbf{#1}}\xspace}
\newcommand{\boldr}{\ensuremath{\mathbf{r}}\xspace}
\newcommand{\xhatb}{\ensuremath{\hat{\mathbf{x}}}\xspace}
\newcommand{\parti}[2]{\frac{ \partial #1}{\partial #2} \xspace}
\newcommand{\partis}[2]{\frac{ \partial^2 #1}{\partial #2^2} \xspace}
\newcommand{\dive}{\ensuremath{\mathbf{\nabla} \cdot}\xspace}
\newcommand{\curl}{\ensuremath{\mathbf{\nabla} \times}\xspace}
\newcommand{\rhat}{\ensuremath{\hat{\mathbf{r}}}\xspace}
\newcommand{\yhat}{\ensuremath{\hat{\mathbf{y}}}\xspace}
\newcommand{\zhat}{\ensuremath{\hat{\mathbf{z}}}\xspace}
\newcommand{\equal}{\ensuremath{&=&}\xspace}
\newcommand{\equals}{\ensuremath{&=&}\xspace}
\newcommand{\psihat}{\ensuremath{\hat{\psi}}\xspace}
\newcommand{\psihatd}{\ensuremath{\hat{\psi}^{\dagger}}\xspace}
\newcommand{\ahat}{\ensuremath{\hat{a}}\xspace}
\newcommand{\Ham}{\ensuremath{\mathcal{H}}\xspace}
\newcommand{\ahatd}{\ensuremath{\hat{a}^{\dagger}}\xspace}
\newcommand{\bhat}{\ensuremath{\hat{b}}\xspace}
\newcommand{\bhatd}{\ensuremath{\hat{b}^{\dagger}}\xspace}
\newcommand{\dr}{\ensuremath{\,d^3\mathbf{r}}\xspace}
\newcommand{\tr}{\ensuremath{\,\mathrm{Tr}}\xspace}
\newcommand{\dk}{\ensuremath{\,d^3\mathbf{k}}\xspace}
\newcommand{\etal}{\emph{et al.\/}\xspace}
\newcommand{\ie}{i.e.}
\newcommand{\fig}[1]{Fig.~\ref{#1}\xspace}
\newcommand{\abs}[1]{\left| #1 \right|}
\newcommand{\Qhat}{\ensuremath{\hat{Q}}\xspace}
\newcommand{\Qhatd}{\ensuremath{\hat{Q}^\dag}\xspace}
\newcommand{\phihatd}{\ensuremath{\hat{\phi}^{\dagger}}\xspace}
\newcommand{\phihat}{\ensuremath{\hat{\phi}}\xspace}
\newcommand{\boldk}{\ensuremath{\mathbf{k}}\xspace}
\newcommand{\bolda}{\ensuremath{\mathbf{a}}\xspace}
\newcommand{\boldx}{\ensuremath{\mathbf{x}}\xspace}
\newcommand{\boldxp}{\ensuremath{\mathbf{x}^{\prime}}\xspace}
\newcommand{\Me}{\ensuremath{\mathbf{M}_e}\xspace}
\newcommand{\boldA}{\ensuremath{\mathbf{A}}\xspace}
\newcommand{\boldp}{\ensuremath{\mathbf{p}}\xspace}
\newcommand{\boldy}{\ensuremath{\mathbf{y}}\xspace}
\newcommand{\boldsigma}{\ensuremath{\boldsymbol\sigma}\xspace}
\newcommand{\boldphi}{\ensuremath{\boldsymbol\phi}\xspace}
\newcommand{\boldalpha}{\ensuremath{\boldsymbol\alpha}\xspace}
\newcommand{\Psihat}{\ensuremath{\hat{\Psi}}\xspace}
\newcommand{\Psihatd}{\ensuremath{\hat{\Psi}^{\dagger}}\xspace}
\newcommand{\Vhatd}{\ensuremath{\hat{V}^{\dagger}}\xspace}
\newcommand{\Xhatd}{\ensuremath{\hat{X}^{\dag}}\xspace}
\newcommand{\Yhatd}{\ensuremath{\hat{Y}^{\dag}}\xspace}
\newcommand{\jhat}{\ensuremath{\hat{J}}\xspace}
\newcommand{\lhat}{\ensuremath{\hat{L}}\xspace}
\newcommand{\Nhat}{\ensuremath{\hat{N}}\xspace}
\newcommand{\rhohat}{\ensuremath{\hat{\rho}}\xspace}
\newcommand{\ddt}{\ensuremath{\frac{d}{dt}}\xspace}
\newcommand{\nset}{\ensuremath{n_1, n_2,\dots, n_k}\xspace}
\newcommand{\Sp}{\ensuremath{S^{\prime}}\xspace}

\newcommand{\divE}{\ensuremath{\mathbf{\nabla \cdot E}}\xspace}
\newcommand{\divA}{\ensuremath{\mathbf{\nabla \cdot A}}\xspace}
\newcommand{\divD}{\ensuremath{\mathbf{\nabla \cdot D}}\xspace}
\newcommand{\divB}{\ensuremath{\mathbf{\nabla \cdot B}}\xspace}
\newcommand{\divj}{\ensuremath{\mathbf{\nabla \cdot j}}\xspace}
\newcommand{\divphi}{\ensuremath{\mathbf{\nabla \cdot \phi}}\xspace}
\newcommand{\curlE}{\ensuremath{\mathbf{\nabla \times E}}\xspace}
\newcommand{\curlH}{\ensuremath{\mathbf{\nabla \times H}}\xspace}
\newcommand{\curlB}{\ensuremath{\mathbf{\nabla \times B}}\xspace}
\newcommand{\curlD}{\ensuremath{\mathbf{\nabla \times D}}\xspace}
\newcommand{\curlJ}{\ensuremath{\mathbf{\nabla \times J}}\xspace}
\newcommand{\curlA}{\ensuremath{\mathbf{\nabla \times A}}\xspace}
\newcommand{\curlphi}{\ensuremath{\mathbf{\nabla \times \phi}}\xspace}
\newcommand{\lapE}{\ensuremath{\mathbf{\nabla^2 E}}\xspace}
\newcommand{\lapB}{\ensuremath{\mathbf{\nabla^2 B}}\xspace}
\newcommand{\lapA}{\ensuremath{\mathbf{\nabla^2 A}}\xspace}
\newcommand{\laphi}{\ensuremath{\nabla^2 \phi}\xspace}
\newcommand{\muj}{\ensuremath{\mathbf{\mu_0 J}}\xspace}

\newcommand{\Hhat}{\ensuremath{\hat{H}}\xspace}
\newcommand{\Vhat}{\ensuremath{\hat{V}}\xspace}
\newcommand{\nhat}{\ensuremath{\hat{n}}\xspace}
\newcommand{\qp}{\ensuremath{{q^\prime}}\xspace}
\newcommand{\chat}{\ensuremath{{\hat{c}}}\xspace}
\newcommand{\chatd}{\ensuremath{{\hat{c}^\dagger}}\xspace}
\newcommand{\boldb}{\ensuremath{\mathbf{b}}\xspace}
\newcommand{\boldq}{\ensuremath{\mathbf{q}}\xspace}
\newcommand{\Htil}{\ensuremath{\tilde{H}}\xspace}
\newcommand{\fdag}{\ensuremath{f_n^\dagger}\xspace}
\newcommand{\fpdag}{\ensuremath{f_{n+1}^\dagger}\xspace}
\newcommand{\fp}{\ensuremath{f_{n+1}}\xspace}
\newcommand{\kp}{\ensuremath{k^\prime}\xspace}
\newcommand{\sk}{\ensuremath{\hat{S}_k}\xspace}
\newcommand{\skdag}{\ensuremath{\hat{S}^\dagger_k}\xspace}
\newcommand{\skp}{\ensuremath{\hat{S}_{k^\prime}}\xspace}
\newcommand{\skpdag}{\ensuremath{\hat{S}^\dagger_{k^\prime}}\xspace}
\newcommand{\sn}{\ensuremath{\hat{S}_n}\xspace}
\newcommand{\sndag}{\ensuremath{\hat{S}^\dagger_n}\xspace}
\newcommand{\sj}{\ensuremath{\hat{S}_j}\xspace}
\newcommand{\sjdag}{\ensuremath{\hat{S}^\dagger_j}\xspace}
\newcommand{\snp}{\ensuremath{\hat{S}_{n+1}}\xspace}
\newcommand{\snpdag}{\ensuremath{\hat{S}^\dagger_{n^\prime}}\xspace}
\newcommand{\zdag}{\ensuremath{\zeta_n^\dagger}\xspace}
\newcommand{\zpdag}{\ensuremath{\zeta_{n+1}^\dagger}\xspace}
\newcommand{\znp}{\ensuremath{\zeta_{n+1}}\xspace}
\newcommand{\zn}{\ensuremath{\zeta_n}\xspace}
\newcommand{\xidag}{\ensuremath{\xi_n^\dagger}\xspace}
\newcommand{\xipdag}{\ensuremath{\xi_{n+1}^\dagger}\xspace}
\newcommand{\xinp}{\ensuremath{\xi_{n+1}}\xspace}
\newcommand{\xin}{\ensuremath{\xi_n}\xspace}

\newcommand{\zmdag}{\ensuremath{\zeta_m^\dagger}\xspace}
\newcommand{\zmpdag}{\ensuremath{\zeta_{m+1}^\dagger}\xspace}
\newcommand{\zmp}{\ensuremath{\zeta_{m+1}}\xspace}
\newcommand{\zm}{\ensuremath{\zeta_m}\xspace}
\newcommand{\ximdag}{\ensuremath{\xi_m^\dagger}\xspace}
\newcommand{\ximpdag}{\ensuremath{\xi_{m+1}^\dagger}\xspace}
\newcommand{\ximp}{\ensuremath{\xi_{m+1}}\xspace}
\newcommand{\xim}{\ensuremath{\xi_m}\xspace}
\newcommand{\fmdag}{\ensuremath{f_m^\dagger}\xspace}
\newcommand{\fmpdag}{\ensuremath{f_{m+1}^\dagger}\xspace}
\newcommand{\fmp}{\ensuremath{f_{m+1}}\xspace}
\newcommand{\fem}{\ensuremath{f_{m}}\xspace}
\newcommand{\Psidk}{\ensuremath{\Psi^\dagger_{k}}\xspace}
\newcommand{\Psik}{\ensuremath{\Psi_{k}}\xspace}
\newcommand{\Psidkp}{\ensuremath{\Psi^\dagger_{k^\prime}}\xspace}
\newcommand{\Psikp}{\ensuremath{\Psi_{k^\prime}}\xspace}
\newcommand{\Psidl}{\ensuremath{\Psi^\dagger_{l}}\xspace}
\newcommand{\Psil}{\ensuremath{\Psi_{l}}\xspace}
\newcommand{\Psidlp}{\ensuremath{\Psi^\dagger_{l^\prime}}\xspace}
\newcommand{\Psilp}{\ensuremath{\Psi_{l^\prime}}\xspace}
\newcommand{\sigp}{\ensuremath{\sigma_{n}^+}\xspace}
\newcommand{\sigm}{\ensuremath{\sigma_{n}^-}\xspace}
\newcommand{\sigmp}{\ensuremath{\sigma_{n+1}^-}\xspace}
\newcommand{\sigpp}{\ensuremath{\sigma_{n+1}^+}\xspace}
\newcommand{\fjdag}{\ensuremath{f_j^\dagger}\xspace}
\newcommand{\fj}{\ensuremath{f_{j}}\xspace}

\newcommand{\Adot}{\ensuremath{\dot{A}}\xspace}
\newcommand{\Adag}{\ensuremath{A^\dagger}\xspace}
\newcommand{\akdag}{\ensuremath{a_k^\dagger}\xspace}
\newcommand{\akdagn}{\ensuremath{a_{k_{N}}^\dagger}\xspace}
\newcommand{\ak}{\ensuremath{a_{k}}\xspace}
\newcommand{\akn}{\ensuremath{a_{-k}}\xspace}
\newcommand{\akndag}{\ensuremath{a_{-k}^\dagger}\xspace}
\newcommand{\bkn}{\ensuremath{b_{-k}}\xspace}
\newcommand{\bkndag}{\ensuremath{b_{-k}^\dagger}\xspace}
\newcommand{\bkdag}{\ensuremath{b_k^\dagger}\xspace}
\newcommand{\bk}{\ensuremath{b_k}\xspace}
\newcommand{\apdag}{\ensuremath{a_p^\dagger}\xspace}
\newcommand{\bpdag}{\ensuremath{b_p^\dagger}\xspace}
\newcommand{\bqdag}{\ensuremath{b_q^\dagger}\xspace}
\newcommand{\bp}{\ensuremath{b_p}\xspace}
\newcommand{\bq}{\ensuremath{b_q}\xspace}
\newcommand{\phidag}{\ensuremath{\phi^\dagger}\xspace}
\newcommand{\phidot}{\ensuremath{\dot{\phi}}\xspace}
\newcommand{\phidotd}{\ensuremath{\dot{\phi}^\dagger}\xspace}
\newcommand{\phiddot}{\ensuremath{\ddot{\phi}}\xspace}
\newcommand{\phiddotd}{\ensuremath{\ddot{\phi}^\dagger}\xspace}
\newcommand{\kn}{\ensuremath{k_{N}}\xspace}
\newcommand{\ako}{\ensuremath{a_{k_1}}\xspace}
\newcommand{\akt}{\ensuremath{a_{k_2}}\xspace}
\newcommand{\akth}{\ensuremath{a_{k_3}}\xspace}
\newcommand{\apo}{\ensuremath{a_{p_1}}\xspace}
\newcommand{\apt}{\ensuremath{a_{p_2}}\xspace}
\newcommand{\akodag}{\ensuremath{a^\dagger_{k_1}}\xspace}
\newcommand{\aktdag}{\ensuremath{a^\dagger_{k_2}}\xspace}
\newcommand{\akthdag}{\ensuremath{a^\dagger_{k_3}}\xspace}
\newcommand{\apodag}{\ensuremath{a^\dagger_{p_1}}\xspace}
\newcommand{\aptdag}{\ensuremath{a^\dagger_{p_2}}\xspace}
\newcommand{\akdago}{\ensuremath{a^\dagger_{k_{1}}^\dagger}\xspace}
\newcommand{\Qdot}{\ensuremath{\dot{Q}}\xspace}
\newcommand{\Qdotd}{\ensuremath{\dot{\hat{Q}}}\xspace}
\newcommand{\aq}{\ensuremath{a_q}\xspace}
\newcommand{\aqdag}{\ensuremath{a_q^\dagger}\xspace}
\newcommand{\lag}{\ensuremath{\mathcal{L}}\xspace}
\newcommand{\pidag}{\ensuremath{\pi^\dagger}\xspace}
\newcommand{\apno}{\ensuremath{a_{p_{N+1}}}\xspace}
\newcommand{\apnodag}{\ensuremath{a^\dagger_{p_{N+1}}}\xspace}
\newcommand{\akno}{\ensuremath{a_{k_{N+1}}}\xspace}
\newcommand{\aknodag}{\ensuremath{a^\dagger_{k_{N+1}}}\xspace}
\newcommand{\apn}{\ensuremath{a_{p_{N}}}\xspace}
\newcommand{\apndag}{\ensuremath{a^\dagger_{p_{N}}}\xspace}
\newcommand{\akf}{\ensuremath{a_{k_{4}}}\xspace}
\newcommand{\akfdag}{\ensuremath{a^\dagger_{k_{4}}}\xspace}
\newcommand{\coef}{\ensuremath{\beta}\xspace}
\newcommand{\akmdag}{\ensuremath{a^\dagger_{k_M}}\xspace}
\newcommand{\apnm}{\ensuremath{a_{p_{N-1}}}\xspace}
\newcommand{\aknd}{\ensuremath{a^\dagger_{k_N}}\xspace}

\newcommand{\apb}{\ensuremath{\hat{a}_{\mathbf{p}}}\xspace}
\newcommand{\apbdag}{\ensuremath{\hat{a}^\dagger_{\mathbf{p}}}\xspace}
\newcommand{\aqbdag}{\ensuremath{\hat{a}^\dagger_{\mathbf{q}}}\xspace}
\newcommand{\Epb}{\ensuremath{E_{\mathbf{p}}}\xspace}
\newcommand{\apbp}{\ensuremath{\hat{a}_{\mathbf{p^{\prime}}}}\xspace}
\newcommand{\apbdagp}{\ensuremath{\hat{a}^\dagger_{\mathbf{p}^{\prime}}}\xspace}
\newcommand{\Epbp}{\ensuremath{E_{\vec{p}^{\prime}}}\xspace}
\newcommand{\boldpp}{\ensuremath{\mathbf{p}^{\prime}}\xspace}
\newcommand{\ppr}{\ensuremath{p^{\prime}}\xspace}
\newcommand{\apbn}{\ensuremath{\hat{a}_{\mathbf{-p}}}\xspace}
\newcommand{\apbdagn}{\ensuremath{\hat{a}^\dagger_{\mathbf{-p}}}\xspace}
\newcommand{\Epbn}{\ensuremath{E_{\mathbf{-p}}}\xspace}
\newcommand{\sighat}{\ensuremath{\hat{\sigma}}\xspace}
\newcommand{\sighatk}{\ensuremath{\hat{\sigma}_k}\xspace}
\newcommand{\phiout}{\ensuremath{\phi_{\mathrm{out}}}\xspace}
\newcommand{\xp}{\ensuremath{x^{\prime}}\xspace}
\newcommand{\boldP}{\ensuremath{\mathbf{P}}\xspace}
\newcommand{\In}{\ensuremath{\mathrm{In}}\xspace}
\newcommand{\Iket}{\ensuremath{\ket{g}_1\ket{g}_2\ket{0}}\xspace}
\newcommand{\pmd}{\ensuremath{\partial_{\mu}}\xspace}
\newcommand{\psis}{\ensuremath{\psi^{*}}\xspace}
\newcommand{\phis}{\ensuremath{\phi^{*}}\xspace}
\newcommand{\psisd}{\ensuremath{\dot{\psi}^{*}}\xspace}
\newcommand{\psid}{\ensuremath{\dot{\psi}}\xspace}
\newcommand{\vy}{\ensuremath{\vec{y}}\xspace}
\newcommand{\apv}{\ensuremath{\hat{a}_{\vec{p}}}\xspace}
\newcommand{\apvdag}{\ensuremath{\hat{a}^\dagger_{\vec{p}}}\xspace}
\newcommand{\aqvdag}{\ensuremath{\hat{a}^\dagger_{\vec{q}}}\xspace}
\newcommand{\Epv}{\ensuremath{E_{\vec{p}}}\xspace}
\newcommand{\apvp}{\ensuremath{\hat{a}_{\vec{p^{\prime}}}}\xspace}
\newcommand{\apvdagp}{\ensuremath{\hat{a}^\dagger_{\vec{p}^{\prime}}}\xspace}
\newcommand{\Epbv}{\ensuremath{E_{\vec{p}^{\prime}}}\xspace}
\newcommand{\vpx}{\ensuremath{\vec{p}\cdot\vec{x}}\xspace}
\newcommand{\vpxy}{\ensuremath{\vec{p}\cdot(\vec{x} -\vec{y}) }\xspace}
\newcommand{\vxy}{\ensuremath{\vec{x}-\vec{y} }\xspace}
\newcommand{\psidag}{\ensuremath{\psi^{\dagger}}\xspace}
\newcommand{\psibar}{\ensuremath{\bar{\psi}}\xspace}
\newcommand{\id}{\ensuremath{\mathds{1}}\xspace}


\newcommand{\phat}{\ensuremath{\hat{p}}\xspace}
\newcommand{\xhat}{\ensuremath{\hat{x}}\xspace}
\newcommand{\Mhat}{\ensuremath{\hat{M}}\xspace}
\newcommand{\Hint}{\ensuremath{\hat{H}_{\mathrm{int}}}\xspace}
\newcommand{\Eint}{\ensuremath{E_{\mathrm{int}}}\xspace}
\newcommand{\mo}{\ensuremath{M_0}\xspace}
\newcommand{\Ea}{\ensuremath{E_{\mathrm{a}}}\xspace}
\newcommand{\Eb}{\ensuremath{E_{\mathrm{b}}}\xspace}
\newcommand{\qhat}{\ensuremath{\hat{q}}\xspace}
\newcommand{\qhatd}{\ensuremath{\hat{q}^\dagger}\xspace}
\newcommand{\ohat}{\ensuremath{\hat{\omega}}\xspace}
\newcommand{\Ahat}{\ensuremath{\hat{A}}\xspace}
\newcommand{\Ahatd}{\ensuremath{\hat{A}^\dagger}\xspace}
\newcommand{\xdot}{\ensuremath{\dot{x}}\xspace}
\newcommand{\tdot}{\ensuremath{\dot{\tau}}\xspace}
\newcommand{\xdoth}{\ensuremath{\hat{\dot{x}}}\xspace}
\newcommand{\hil}{\ensuremath{\mathcal{H}}\xspace}
\newcommand{\hilint}{\ensuremath{\mathcal{H}_\mathrm{int}}\xspace}
\newcommand{\hilex}{\ensuremath{\mathcal{H}_\mathrm{ext}}\xspace}
\newcommand{\po}{\ensuremath{\hat{p}_0}\xspace}
\newcommand{\Atil}{\ensuremath{\tilde{A}}\xspace}
\newcommand{\Atild}{\ensuremath{\tilde{A}^\dagger}\xspace}
\newcommand{\Mtil}{\ensuremath{\tilde{M}}\xspace}
\newcommand{\otil}{\ensuremath{\tilde{\omega}}\xspace}
\newcommand{\ktil}{\ensuremath{\tilde{k}}\xspace}
\newcommand{\Sq}{\ensuremath{\hat{S}(\zeta)}\xspace}
\newcommand{\pr}{\ensuremath{\hat{\Pi}}\xspace}
\newcommand{\prz}{\ensuremath{\hat{\Pi}_0}\xspace}
\newcommand{\pro}{\ensuremath{\hat{\Pi}_1}\xspace}
\newcommand{\atilo}{\ensuremath{\tilde{a}_0}\xspace}
\newcommand{\atildo}{\ensuremath{\tilde{a}^\dagger_0}\xspace}
\newcommand{\atilon}{\ensuremath{\tilde{a}_1}\xspace}
\newcommand{\atildon}{\ensuremath{\tilde{a}^\dagger_1}\xspace}
\newcommand{\Htilo}{\ensuremath{\tilde{H}_0}\xspace}
\newcommand{\Stil}{\ensuremath{\tilde{S}}\xspace}

\newcommand{\heff}{\ensuremath{\h H_{\mathrm{eff}}}\xspace}


\author{Rebecca Haustein}
\author{Gerard J.~Milburn}
\author{Magdalena Zych}
\affiliation{Centre for Engineered Quantum Systems, School of Mathematics and Physics, The University of Queensland, St Lucia, QLD 4072, Australia }

\title{Mass-energy equivalence in harmonically trapped particles}

\begin{abstract}
Precise understanding of the dynamics of trapped particles is crucial for nascent quantum technologies, including atomic clocks and quantum simulators. Here we present a framework to  systematically include quantum effects arising from the mass-energy equivalence in harmonically trapped systems. We find that the mass-energy equivalence leads to squeezing, displacement and frequency changes of harmonic modes associated with different internal energies. The framework predicts new phenomena, notably, the existence of a lower bound to the fractional frequency shift in atomic clocks arising from the interplay between gravitational effects and so-called time dilation shifts. Analogous effects will arise in other trapping potentials, especially in periodic lattices, and may play a role in correlation dynamics and thermalisation process in many-body systems and cold gases.\end{abstract}

\maketitle
\textit{Introduction.--} 
Trapped atoms find applications in the most stable atomic clocks \cite{RevModPhys.87.637},  quantum information processing~\cite{HAFFNER2008155, saffman2010quantum, Ciaramicoli2003},  simulations of high-energy physics and cosmology \cite{TrappedIons_Schneider:2012}, in tests of fundamental physics~\cite{safronova2018search, katori2011optical} including  precision measurements of the Standard Model \cite{Hanneke2008, Odom2006, Gabrielse2006}, and novel experiments studying thermalisation processes in  
closed-systems~\cite{Gogolin_2016, Erne:420237}. 

It is well understood,  already in classical-physics, that relativistic mass-energy equivalence -- specifically, that changing the internal energy of a system also changes its mass --  is equivalent to introducing relativistic time dilation \cite{Einstein:1905, Einstein:1907, PhysRevLett.4.341, Greenberger:1970_1, Greenberger:1970_2, Greenberger:1974}. Recently, new effects were found  when taking into consideration that the internal energy contribution to the mass 
requires a quantum description.  For free particles it was shown that quantised mass-energy and the associated gravitational and special relativistic time dilation lead to novel effects in interference of ``quantum clocks'' \cite{Zych:2011, Zych:2012, bushev2016single} and generic composite particles~\cite{pikovskiuniversal2015, PikovskiTime2017, korbicz2017information}, which have been further studied in double-slit type experiments in a gravitational field~\cite{2016PhRvL.117i0401P, Orlando2017}. Limitations to the notion of an ideal clock arising from the associated phenomena have also been explored for free and trapped particles~\cite{paige2018quantum, Sinha_2014}, and for interacting quantum clocks~\cite{Castro:2017clocks}. Furthermore, for atoms interacting with radiation, quantised mass-energy removes an alleged ``friction''~\cite{sonnleitner2017will} due to spontaneous emission, resolving an apparent tension between the relativistic treatment of radiation and  the non-relativistic treatment of atoms~\cite{sonnleitner2018mass}. Indeed, the symmetry group of a system whose centre of mass (CM) is non-relativistic but which includes the mass-energy equivalence is not a Galilei group, but its non-trivial central extension~\cite{ZychGreenberger2019}.

For trapped particles, quantised mass-energy equivalence results in an energy spectrum that is not a  simple sum of the internal and CM energies -- since  generically the CM energy in an external potential depends on the particle's mass. This was explored in a toy-model of a trapping potential -- an infinite square-well~\cite{krause2016taking} -- and to propose a realisation of quantum clock interference with trapped electrons~\cite{bushev2016single}. In the latter, the internal energy was associated with electronic spin -- which is only analogous to the rest mass-energy at lowest relativistic order (as the latter is a relativistic scalar). A full quantised dynamics of trapped particles with dynamical mass-energy has not been developed.

In this work we derive a framework for incorporating mass-energy equivalence in generic harmonically trapped quantum particles and discuss the associated physical effects.
We trace back the so-called ``time dilation shifts'' already measured in atomic clocks~\cite{PhysRevLett.104.070802, PhysRevLett.116.063001, PhysRevLett.118.053002} -- and associated therein to secular motion of the trapped particle -- to the semi-classical limit of the mass-energy effects, which is consistent with the approximate treatment in~\cite{paige2018quantum, Yudin:2018}. Crucially, our approach  reveals further physical effects from quantised mass-energy, that to our knowledge have not been previously discussed. 
 
\textit{Harmonically trapped particles with quantised internal energy.--} 
Physical states of an effectively point-like particle with quantised internal energy can be described in a tensor product Hilbert space $\cH_{int}\otimes\cH_{cm}$, where $\cH_{int}$ and $\cH_{cm}$ refer to the internal and CM degrees of freedom (DOFs), respectively. A free Hamiltonian of the  particle on a static symmetric space-time with metric $g_{\mu\nu}$ and signature $(-, +,+,+)$ reads~\cite{zych2015PhD, PikovskiTime2017, zych2015quantumEEP, Anastopoulos:2018, zych2018gravitational}  
$\sqrt{-\go(c^2\h p_j\h p^j +\h M^2 c^4)}$ where $\h p_j$, $j=1,2,3$,  is the canonical CM momentum and $\h Mc^2=M_0c^2\h I+\hint$ is the total mass-energy, which includes the rest mass $M_0$  as well the internal energy operator $\hint$. 
 At low energies the relativistic Hamiltonian reduces to  $Mc^2+{p^2}/{2M}+M\phi(x)$, with $\phi(x)$ denoting the gravitational potential. The split of the relativistic mass-energy 
into mass and energy is a-priori arbitrary; the choice above is such that the lowest eigenvalue of $\hint$ is $E_0=0$ and  so $M_0c^2$ is the mass-energy associated with the internal ground state.  For systems such as atoms, ions or molecules $\hint$ describes the electronic or vibrational energy levels and in the operator norm satisfies $|| \hint/M_0c^2||\ll1$. To lowest order in $1/c^{2}$ the Hamiltonian then reads $H_{int}(1-{p^2}/{2M_0c^2}+\phi(x)/c^2)+{p^2}/{2M_0}+M_0\phi(x)$.  The first term describes the well-understood and directly measured special and general-relativistic time dilations of the internal dynamics~\cite{Wineland:2010, Reinhardt2007test} --  and directly demonstrates the above-mentioned relation between time dilation and mass-energy equivalence. 
We now consider a low-energy quantum particle in a harmonic potential of stiffness $k$ and a homogeneous gravitational field with acceleration $g$,  described by the Hamiltonian
\be
H =  \Mhat c^2+\frac{\phat^2}{2\Mhat} +\h M g\h x+ \frac{1}{2}k\xhat^2.
\ee{Ham}
\eq{Ham} applies in a regime where the CM is effectively non-relativistic  but where the internal evolution is fast enough to be sensitive to time dilation. This is well suited for describing trapped atom or ion experiments.

Since internal energy commutes with the CM operators $\h x$, and $\h p$, the eigenstates of $\h H$ are products of  internal eigenstates $\ket{E_i}$, such that $\h M\ket{E_i}= M_i\ket{E_i}$ where $M_i:=M_0+E_i/c^2$, $i\in\mathbb{N}$ and the CM eigenstates -- which we derive below. Introducing orthonormal projectors on the internal energies,  $\h\Pi_i:=\proj{E_i}$, \eq{Ham} takes the form $\sum_i \h{h}_i\otimes\h\Pi_i$ where 
$\h h_i = M_ic^2 + {\phat^2}/{2M_i} + k\xhat^2/2+M_ig\h x$.
For each internal energy  we have a harmonic oscillator (HO) with mass $M_i$,  frequency $\omega_i:=(k/M_i)^{1/2}$ and spatial coordinate $\h x_{g,i}:=\h x+g/\omega_i^2$. Since $[\h x_{g,i}, \h p]=i\hbar$ the normal mode operators for each $M_i$ read
\be
\h {a}_{i}\!=\! \sqrt{\frac{M_i\omega_i}{2\hbar}} \big(\xhat_{g,i} + \frac{i\,\phat}{M_i \omega_i}\big).
\ee{aM}
Using $\h n_{i}=\h a_{i}^\dagger\h a_{i}$, the operators $\h h_i$ are diagonalised as usual
\be
\h h_i=M_ic^2(1-\frac{g^2}{2\omega_i^2c^2})+\hbar\omega_i(\h n_{i}+\frac 1 2).
\ee{ham_i}
For each internal energy the CM eigenstates are  Fock states $\sqrt{n!}\ket{n_{i}}=\h a_{i}^{\dagger n}\ket{0_i}$ where $\ket{0_i}$ is the CM ground state associated with the $i^{th}$ internal energy. The eigenbasis of $\h H$ is thus of a product form $\{\ket{E_i}\ket{n_{i}}\}_{i, n_{i}\in\mathbb{N}}$ but is not a product of the two bases. In particular, the CM modes associated with an $i^{th}$ internal state and with the internal ground state are related by squeezing and displacement
\be
\h a_{i} =S^\dagger(r_i)D^\dagger(\alpha_{gi})\h a_{0}D(\alpha_{gi})S(r_i),
\ee{squeezing}
where $ S(r_i)=e^{r_i(\h a_{0}^2-\h a_{0}^{\dagger2})/2}$ and $D(\alpha_{gi})=e^{\alpha_{gi}(\h a_{0}^\dagger - \h a_{0})}$ with
$2\cosh(r_i)=(\frac{M_0}{M_i})^{\frac1 4}+(\frac{M_i}{M_0})^{\frac 1 4}$,  $2\sinh(r_i)=(\frac{M_0}{M_i})^{\frac1 4}-(\frac{M_i}{M_0})^{\frac 1 4}$,  $\alpha_{gi}=\frac{g\Delta M_i}{\sqrt{2\hbar M_i\omega_i^3}}$,  $\Delta M_i = M_i-M_0\equiv E_i/c^2$.  \eq{squeezing} equivalently reads $\h a_{i} = \cosh(r_i)\h a_{0}-\sinh(r_i)\h a_{0}^\dagger+\alpha_{gi}$. We note that to lowest order in $\Delta M_i$ the displacement reads
\be
\alpha_{gi}\approx g\frac{\Delta M_i}{\sqrt{2\hbar M_0\omega_0^3}},
\ee{alpha_g}
and the CM frequencies associated with the excited and the ground state relate as  \be
\frac{\omega_i}{\omega_0}=\sqrt{\frac{M_0}{M_i}}\approx1-\frac{\Delta M_{i}}{2M_0}.
\ee{freq_shift}
The frequency change is a classical effect in the sense that the same relation holds for two classical harmonic oscillators with masses $M_0, M_i$. 

\textit{Physical effects from mass-energy equivalence in trapped particles.--} 
Let us consider a pair of  internal states  where for each of them them the CM is in the $n^{th}$ Fock state, namely $\ket{E_i}\ket{n_{i}}$, and $\ket{E_0}\ket{n_{0}}$. Their total energy difference, obtained from \eq{ham_i}, is $
M_ic^2(1-\frac{g^2}{2\omega_i^2c^2}) - M_0c^2(1-\frac{g^2}{2\omega_0^2c^2})+\hbar(\omega_i-\omega_0)(n+\frac{1}{2})$ and to lowest order in $1/c^2$ it reads 
\be
E_i\left(1-\frac{g^2}{\omega_0^2c^2}-\frac{\hbar\omega_0}{2M_0c^2}\left(n+\frac{1}{2}\right)\right),
\ee{energy_gap}
where we have used $M_i^2-M_0^2\approx\Delta M_i2M_0$. Neglecting the gravitational effects and the mode difference (i.e.~$\h n_i\equiv \h n$ for all i) \eq{energy_gap} entails a fractional frequency shift $\frac{-\hbar\omega_0}{2M_0c^2}(\mean{\h n}+\frac1 2))$. This is known in atomic clocks as time dilation shift: the CM energy  $\hbar\omega_0(\mean{\h n}+\frac1 2)$ is interpreted as twice the kinetic energy $M_0\mean{v^2}$; whereby the shift is given by $\mean{v^2}/2c^2$ and is interpreted as due to special-relativistic time dilation caused by the CM ``motion'' \cite{PhysRevLett.104.070802, PhysRevLett.116.063001, PhysRevLett.118.053002}. For atomic mass $M_0\!\sim\!10^{-26}$kg trap frequency $\omega_0\!\sim\!1$MHz and $\mean{\h n}\!\sim\!1$ our approach gives a shift of magnitude $10^{-19}$, in full agreement with the measured values \cite{brewer201927}. Unless otherwise stated all numerical estimations  henceforth in this article are for the above values of $M_0$ and $\omega_0$.
By cooling the CM motion the time dilation shift can only be suppressed to $\frac{\hbar\omega_0}{4M_0c^2}$ which is of the order $10^{-20}-10^{-19}$.  Indeed, uncertainty in this shift is a dominant contribution to the overall uncertainty of state-of-the-art clocks based on trapped ions \cite{PhysRevLett.116.063001, brewer201927}. 

Importantly,   \eq{energy_gap} entails that the fractional frequency shift due to mass-energy equivalence 
has an absolute minimum 
\be
\delta_{min}=-\frac{3}{2\sqrt[3]{2}}\left(\frac{\hbar g (n+\frac{1}{2})}{c^3M_0}\right)^{\frac{2}{3}}
\ee{min_frac} for $ \omega_{0}^{min}(n)=\big(\frac{4g^2M_0}{\hbar(n+{1}/{2})}\big)^{\frac{1}{3}}$. 
It is lowest at $n=0$, $\omega_{0}^{min}(0)\!\sim\!4$kHz, and takes the value $\delta_{min}\!\sim\!10^{-22}$ -- just two orders of magnitude below the minimal time-dilation shift -- hence it may be relevant for next generation clocks. 
We stress that \eq{energy_gap} is the exact energy difference between two eigenstates of the Hamiltonian in \eq{Ham}, and describes a shift -- rather than uncertainty~\cite{Sinha_2014} -- in the transition frequency~\footnote{In ref.~\cite{Sinha_2014} an intuitive, semiclassical picture was used to analyse frequency uncertainties in trapped atoms based on the assumption that there exist a joint ground state of the CM for all internal states, and in addition interpreting semi-classically the position and ``acceleration'' uncertainties of the atom.}. This shows that including the full description of the CM modes can allow us to find optimal states for mitigating uncertainties in transition frequencies in trapped atoms.

\begin{figure}[h!]
\includegraphics[width=0.9\columnwidth]{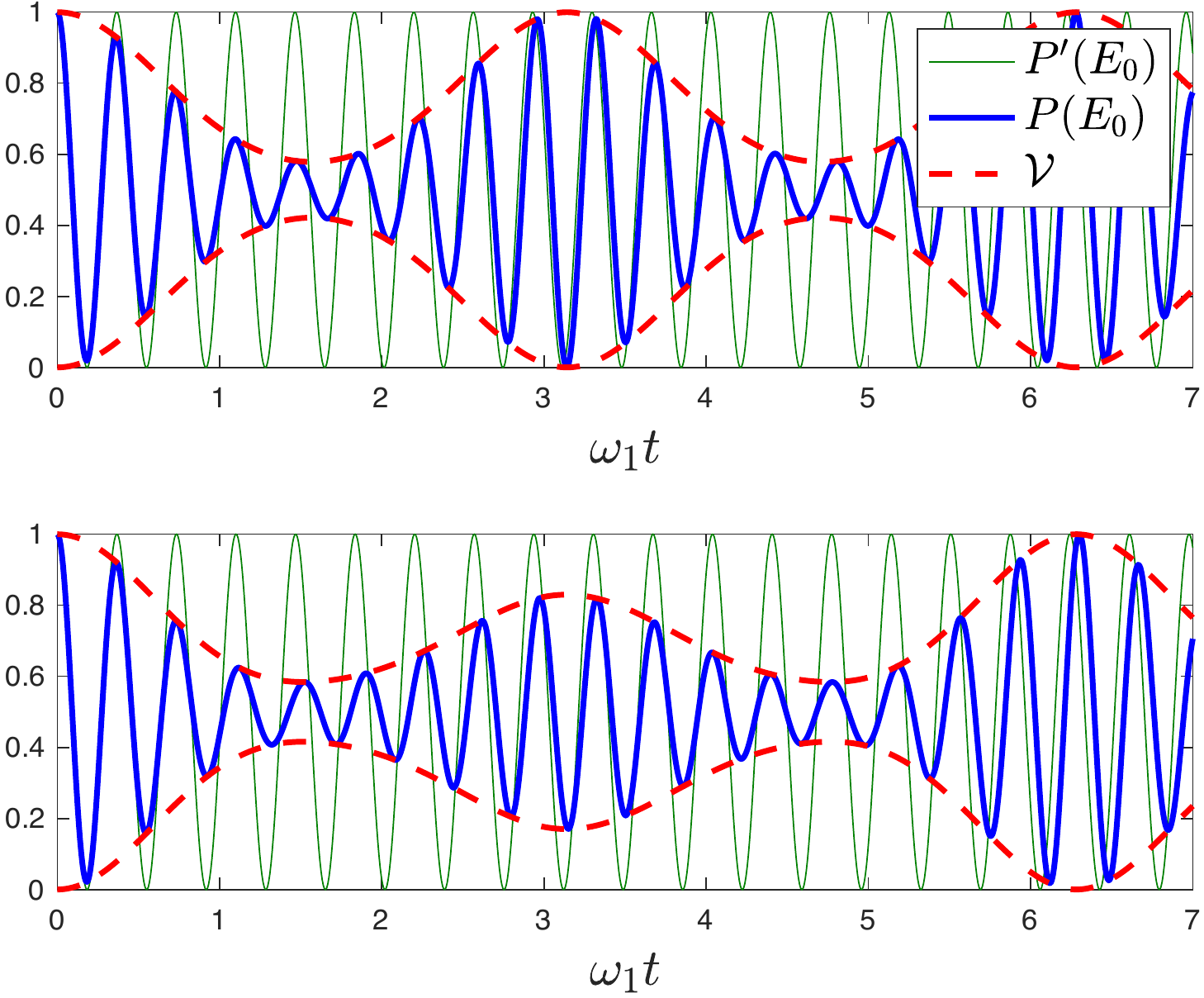}
\caption{Interference described by \eq{probab} 
for an initial number state $n_0=10$ and $\Delta M/M_0=0.5 $  top: without gravity, bottom: with gravity. Thick blue line is the full interference $P_{E_0}$,  the associated visibility  is given be the dashed red line; the thin green line shows interference arising from a non-relativistic theory, where only a phase is acquired, $P'(E_0)=\frac{1}{2}(1+\cos(\omega_ct))$ where $\omega_c:=E_1/\hbar$. The visibility reaches minimum at $t_{\mathrm{min}}\approx\pi/2\omega_1$; For the initial vacuum $\mathcal{V}(t_\mathrm{min})= e^{\frac{-a_0 x_0^2}{1 + S^2}} \sqrt{\frac{2S}{1 + S^2}}$ where $S=\sqrt{M_0/M_1}$, $a_0=M_0\omega_0/\hbar$, $x_0=g/\omega_0^2$ and becomes smaller for higher number states. The partial revival occurs at $t_\mathrm{rev}=2t_\mathrm{min}2$ where $\mathcal{V}(t_{\mathrm{rev}})= e^{-a_0 x_0^2}$. }
\label{Figure1}
\end{figure}

We now proceed to discuss the quantum effects of the  mass-energy equivalence associated with \eq{Ham} in  the context of Ramsey spectroscopy. %
Since mass-energy equivalence induces an interaction between the internal and CM DOF via Eqs.~(\ref{ham_i}--\ref{freq_shift}), it  can entangle CM and internal states and affect time evolution even for an initial product state.  In particular, the time-evolved CM state $\h \rho_{\mathrm{cm}}(t)$ is in general mixed and depends non-trivially on the initial internal state $\h\rho_{int}$ through $\h \rho_{\mathrm{cm}}(t)=\Tr_{\mathrm{int}}\{\h U(t)\h\rho_{cm}(0)\h\rho_{int}\h U^\dagger(t)\} = \sum_{k}p_k \h U_k(t)\h\rho_{cm}(0)\h U_k^\dagger(t)$, where $\h U_i(t):=e^{-i\h h_i t/\hbar}$ and $p_k=\bra{E_k}\h\rho_{int}\ket{E_k}$ is the internal energy distribution in $\h\rho_{int}$. 
To investigate the phenomenology associated with these effects in Ramsey spectroscopy,  consider internal DOF of a trapped particle prepared in the ground state $\ket{E_0}$ while its CM in a generic,  possibly mixed, state~\footnote{The subscript 0 refers to the fact that this CM state has been prepared with the internal DOF in the ground state.} 
${\h\rho_0}$. A $\pi/2$-pulse takes the internal state into a superposition $(\ket{E_0}+\ket{E_1})$ and after time $t$ another $\pi/2$-pulse is applied. The probability to detect the particle in the internal ground state immediately after this sequence is 
\be
P({E_0})=\frac{1}{2}+\frac{1}{4}\left(\Tr\{\h U_1(t)\h \rho_0\h U_0^\dagger(t)\}+c.c\right),
\ee{probab}
where the trace is over the CM, and ``c.c'' stands for the complex conjugate of the preceding term. 
%
 For a pure initial state $\h \rho_0=\ket{\psi_0}\bra{\psi_0}$ the trace reads $\bra{\psi_0}\h U_0(t)^\dagger S^\dagger(r)D^\dagger(\alpha_g)\h U_0(\omega_1,t)D(\alpha_{g})S(r)\ket{\psi_0}$, where $\h U_0(\omega_1, t)$ means that $\h n_{1}$ in $\h h_1$  is already replaced by $\h n_0$ using \eq{squeezing}, and  
$r\equiv r_1$, $\alpha_{g,1}\equiv\alpha_g$. 
The trace quantifies 
the extent to which the joint action of the displacement and squeezing commutes with time evolution. In \fig{Figure1} we plot numerical results for an initinal number state, with and without gravity\footnote{Physically, neglecting gravity corresponds to looking at horizontal modes of  trapped particles, so that the dynamics in the direction of gravitational acceleration factors out.}. The latter case is obtained by setting $g=0$ (and \eq{squeezing} reduces to pure squeezing).  Apart from a shift in the frequency of the internal transition the fringe visibility is modulated, with a full revival at $t=\pi/\omega_1$ in the gravity-free case and $t=2\pi/\omega_1$ when gravity is included (with a partial revival at $\pi/\omega_1$). For higher number states or larger $\Delta M/M_0$ the minimum of the visibility deceases.  For coherent states the results exhibit the same general  features and have a simple analytical form given in Supplementary Material.

Below we discuss general features of \eq{probab} in the approximation of small displacements and squeezing, which is justified for the previously stated parameters and an optical energy gap between the internal states $\omega_c:=E_1/\hbar\!\sim\!10^{15}$Hz,  where $r\!\sim\!10^{-10}$ and $\alpha_g\!\sim\!10^{-15}$. 
Under this approximation $U_0^\dagger\h U_1\approx e^{-i(\omega_c+\heff) t}$ where $\heff:=-\omega_c{g^2}/{\omega_0c^2}+\omega_1(\h n_1+1/2)-\omega_0(\h n_0 +1/2)+it\omega_0\omega_1/2 [\h n_0, \h n_1]$, and 
\be
P({E_0})=\frac{1}{2}\left\{1+\mathcal{V}\cos\left((\omega_c+\langle{\heff}\rangle) t\right)\right\};
\ee{probab_approx}
  the mean $\mean{\cdot}$ is taken with respect to $\h\rho_0 $ and $\mathcal{V}\equiv |\Tr\{\h U_1\h \rho_0\h U_0^\dagger\}|$, explicitly
$\mathcal{V}\!\approx\!{1\!-\!(\Delta\h H_{\mathrm{eff}}/\sqrt{2}\hbar)^2}$ where
$\Delta\h H_{\mathrm{eff}}^2=\langle\h H_{\mathrm{eff}}^2\rangle-\langle\h H_{\mathrm{eff}}\rangle^2$.  
 The mass-energy effects thus induce a frequency shift given by the mean effective energy and visibility modulations given by the effective energy variance. 
The frequency shift reads 
\be
\begin{split}
 \langle{\heff}\rangle =&-\omega_c\frac{g^2}{\omega_0c^2}+ \Delta\omega\big(\langle\h n_0\rangle+\frac1 2\big) \\+ &\omega_1\langle\h n_1-\h n_0\rangle+\frac{it}{2}\omega_0\omega_1\langle[\h n_0, \h n_1]\rangle,
\end{split}
\ee{full_shift}
where $\Delta\omega:=\omega_1-\omega_0\approx{-\omega_0\hbar\omega_c}/{2M_0c^2}$. The top line of \eq{full_shift} includes the semi-classical effects due to the mass change of the oscillator, specifically, the second term is the time dilation shift seen in atomic clocks (cf \eq{energy_gap} and discussion below). 
The two terms in the bottom line are solely due to the CM mode change; using \eq{squeezing} we find that they are a sum of two terms $A_0+\alpha_g\omega_1A_g$, the gravity free $A_0= \omega_1\sinh^2(r)-\frac{\omega_1}{2}\sinh({2r})\left(\langle\h a^{2}_0\rangle(1-it\omega_0) + \langle \h a^{\dagger2}_0\rangle(1+it\omega_0) \right)$; and $A_g=e^{-r}(1+\frac{it\omega_0}{2})\langle{\h a^\dagger_0}\rangle+e^{-r}(1-\frac{it\omega_0}{2})\mean{\h a_0}+\alpha_g$.
For any state diagonal in the Fock basis the visibility reads
\be
\mathcal{V} \approx 1-\frac{1}{2}\left(t\Delta[\omega_1\h n_1-\omega_0\h n_0]\right)^2,
\ee{visib_approx} 
where $\Delta[\omega_0\h n_0-\omega_1\h n_1]$ is the  squared variance of $\omega_0\h n_0-\omega_1\h n_1$.
Neglecting the squeezing for clarity we obtain $\omega_0\h n_0-\omega_1\h n_1\approx \Delta\omega n_0 +\omega_1\alpha_g(\h a_0^\dagger +\h a_0)+\omega_1\alpha_g^2$ which yields $\mathcal{V}_{\mathrm{f}}\approx 1-(\omega_0\Delta\h n_0t)^2 -(\omega_1\alpha_gt)^2(2\mean{\h n_0} +1)$. Specialising to a thermal state at temperature~$T\!\sim\!1$mK we have $\Delta\h n_0\approx \mean{\h n_0}\approx k_BT/\hbar\omega_0\!\sim\!10^2$ and
the first, gravity independent, term reads $\omega_c\frac{k_BT}{2M_0c^2}\!\sim\!1$mHz, while the gravity dependent term reads $\frac{g\omega_c}{\omega_0c^2}\sqrt{\frac{k_BT}{M_0}}\!\sim\!10^{-7}$Hz. We note that the corresponding time dilation shift is ${k_BT}/{2M_0c^2}\!\sim\!10^{-18}$, in  agreement with experiments~\cite{PhysRevLett.116.063001}. The modulation of the visibility arises due to quantum correlations developed between the CM and the internal states, and is therefore of lower order  than the corresponding frequency shift. Furthermore, the gravity-dependent effect of  the mass-energy equivalence is, at least for thermal states, several  orders of magnitude smaller than the inertial effects.

Below we show that the CM mode squeezing~\eq{squeezing} can become amplified under periodic driving.  Consider the particle initially in the  eigenstate $\ket{E_0, n_0}$; the internal state at $t=0$ is excited to $\ket{E_1}$ and then periodically de-excited to $\ket{E_0}$ and excited back to $\ket{E_1}$ at time intervals $t_i=\pi/2\omega_i$, $i=1,0$. Each iteration results in squeezing and displacement: $\h U_0(t_0)\h U_1(t_1)= -S(2r)D(\beta_g(r))$ where $\beta_g(r):=\alpha_g(\cosh(r)(1-i)-\sinh(r)(1+i))$. However, repeating the process again, to linear order in $\Delta M/M_0$, gives pure squeezing.  Thus, for an even number $N$ of the iterations, the effective transformation resulting from such a periodic process is squeezing given by $S(2Nr)$.  To lowest order in $1/c^2$ the effective squeezing parameter reads $2Nr\approx N\frac{E_1}{2M_0c^2}$, and for an optical energy gap is $\!\sim\!N10^{-10}$. This means, for example, that the position variance of an initial state will increase by $1$\% after $n\sim10^8$ iterations, and a total time of $\!\sim\!100$s. 
We note that for the initial vacuum state, overlap with the final state $P_N=|\bra{\psi_0}S(2Nr)\ket{\psi_0}|^2$ exhibits a simple Gaussian decay with $N$, however, for a number state it can exhibit   revivals~\cite{schleich1987oscillations, KimKnight:1989PRA}. For a direct observation of this amplified squeezing one can use parameter estimation techniques from quantum metrology~\cite{toth2014quantum, Paris:2009Qest, NolanHaine:2017PRA} and explore to which extent the optimal strategies for this task already derived in optics~\cite{Milburn:1994Squeeze, Chiribella:2006Squeezing} can be physical realised in the present context. 

Finally let us note that even for an initial product-state simply evolving under \eq{Ham}, the mass-energy effects can become significant. As an illustration consider CM initially in a coherent state $\ket{\alpha}$ w.r.t to mode $\h a_0$, such that $\h a_0\ket{\alpha}=\alpha\ket{\alpha}$.
The $Q$-function of the time evolved state $Q(\beta)=\bra{\beta}\h \rho_{\mathrm{cm}}(t)\ket{\beta}$ 
 at short times is 
\be
Q(\beta)\approx|\!\braket{\beta| \alpha}\!|^2\sum_kp_ke^{-(\omega_kt)^2\big(\frac{\bra{\beta}\h n_k^2\ket{\alpha}}{\braket{\beta| \alpha}}-\big(\frac{\bra{\beta}\h n_k\ket{\alpha}}{\braket{\beta|\alpha}}\big)^2\big)}.
\ee{Qfunct}
For $\alpha=0$, up to $\mathcal{O}(1/c^2)$, $Q(\beta)\approx \e{-\abs{\beta}^2}(1-(\omega_0t)^2\frac{\mean{H_{int}}}{2M_0c^2}(\abs{\beta}^2+\beta g\sqrt{\frac{2M_0}{\hbar\omega_0^3}}))$, where $\beta\in\mathbb{R}$ for simplicity. Neglecting the term $ g\sqrt{{2M_0}/{\hbar\omega_0^3}}\!\sim\!10^{-5}$ gives 
$Q(\beta) \approx \e{-\abs{\beta}^2(1+\frac{(\omega_0t)^2\mean{H_{int}}}{2M_0c^2})}$ which means that the vacuum becomes squeezed, with the effective squeezing parameter $r_{\mathrm{eff}}=(\omega_0t)^2{\mean{H_{int}}}/{2M_0c^2}$. For the trapped system comprising $N\gg1$ atoms of  mass $m_0c^2$ and approximating internal states as harmonic modes, in a high temperature limit  $\mean {H_{int}}\approx 3N k_B T$. Taking $T\!\sim\!100$K and $m_0\!\sim\!10^{-26}$kg 
we have $\mean{H_{int}}/2M_0c^2\!\sim\!10^{-12}$ and $r_{\mathrm{eff}}\!\sim\! 1$ on the time scale of $1$s. Thus even for the most robust state, the vacuum, the mass-energy effects can become significant.

\textit{Discussion.--} 
Our results show that the relativistic mass-energy equivalence induces new effects in trapped systems --  squeezing and displacement -- in addition to currently conisdered frequency shifts. These effects will be relevant for precision metrology and tests of fundamental physics and can be of high interest for quantum thermodynamics. We expect them to become more significant for interacting many body systems in periodic lattices such as in trapped-ion quantum simulators where the build-up of quantum correlations and precise control of the dynamics are of  interest~\cite{Gaertner:2017OTOC, LewisSwan:2019FOTOC}. 

The time dilation shift already limits the precision of atomic clocks to~\cite{PhysRevLett.116.063001, brewer201927} $10^{-19}$.  A more accurate description of the mass-energy effects will therefore be required to reach the desired precision of $10^{-20}$ in next-generation clocks. Moreover, including the gravitational acceleration we found an absolute minimum to  the fractional shift of internal frequency at the level of  $10^{-22}$. To fully analyse the implications of these effects in atomic clocks  additional atom-laser interactions will be incorporated as a next step. As an aside, we note that if the trap stiffness $k$ was made internal energy dependent -- so that the CM frequency, rather than the potential~\cite{katori2011optical}, is the same for the relevant clock-states -- the squeezing effect will still be present, with twice as large squeezing parameter. 

For test of fundamental physics, the mass-energy induced interactions between CM and internal states are relevant to the proposed tests of quantum effects from time dilation on quantum interference~\cite{Zych:2011, pikovskiuniversal2015, roura2018gravitational, bushev2016single} and  tests of the Einstein Equivalence Principle for quantum tests masses~\cite{zych2015quantumEEP, 2016CQGra..33sLT01O, geiger2018proposal, Anastopoulos:2018}, where the dynamics associated with the internal states plays a key role~\cite{zych2015quantumEEP, rosi2017quantum}.

In the context of thermodynamics at the quantum scale~\cite{Anders_2017, binder2018thermodynamics} the mass-energy equivalence may play a role in thermalisation processes in trapped systems even at  moderate conditions where relativistic effects would otherwise not be expected, cf~\eq{Qfunct}.  More generally, the very question of thermalisation becomes non-trivial in the present context. A thermal state of a trapped system, for clarity setting $g=0$, reads $\rho_T=\frac{1}{\mathcal{Z}}\sum_{k}e^{-\beta\h{h}_k}\!\otimes\!\h\Pi_k$, where the sum is over the internal states and  $\mathcal{Z}= \sum_ke^{-\beta (E_k+\hbar\omega_k/2)}/(1-e^{-\beta\hbar\omega_k})$. It is not a product of thermal states as each internal state is correlated with a squeezed thermal state of the CM. It is therefore conceivable that the mass-energy effects may play a role in yet to be fully understood thermalisation precesses in isolated systems \cite{Erne:420237, Gogolin_2016}.  
 
Finally, we note that higher order relativistic corrections to the CM, such as $\propto p^2/M_0^3c^2$ can be incorporated into our model without a conceptual difficulty, allowing to extend the treatment to higher CM energies.  

\textit{Acknowledgment--}
We thank Fabio Costa, Simon Heine, Peter Knight, Jon Links, David Wineland, and Jun Ye for discussions. R.H., G.J.M and M.Z.~acknowledge the Australian Research Council (ARC) Centre of Excellence for Engineered Quantum Systems (EQuS) CE170100009; M.Z.~acknowledges the ARC grant DECRA  DE180101443, the University of Queensland (UQ) Early Career Researcher Grant UQECR1946529; and the traditional owners of the land on which UQ is situated, the Turrbal and Jagera people.

\bibliographystyle{linksen}
\bibliography{bibliotrapped}

\providecommand{\href}[2]{#2}\begingroup\raggedright\begin{thebibliography}{10}

\bibitem{RevModPhys.87.637}
A.~D. Ludlow, M.~M. Boyd, J.~Ye, E.~Peik, and P.~O. Schmidt, ``Optical atomic
  clocks,'' \href{http://dx.doi.org/10.1103/RevModPhys.87.637}{{\em Rev. Mod.
  Phys.} {\bfseries 87}, 637--701 (2015)}.

\bibitem{HAFFNER2008155}
H.~H{\"a}ffner, C.~Roos, and R.~Blatt, ``Quantum computing with trapped ions,''
  \href{http://dx.doi.org/https://doi.org/10.1016/j.physrep.2008.09.003}{{\em
  Physics Reports} {\bfseries 469}, 155 -- 203 (2008)}.

\bibitem{saffman2010quantum}
M.~Saffman, T.~G. Walker, and K.~M\o{}lmer, ``Quantum information with Rydberg
  atoms,'' \href{http://dx.doi.org/10.1103/RevModPhys.82.2313}{{\em Rev. Mod.
  Phys.} {\bfseries 82}, 2313--2363 (2010)}.

\bibitem{Ciaramicoli2003}
G.~Ciaramicoli, I.~Marzoli, and P.~Tombesi, ``Scalable Quantum Processor with
  Trapped Electrons,''
  \href{http://dx.doi.org/10.1103/PhysRevLett.91.017901}{{\em Phys. Rev. Lett.}
  {\bfseries 91}, 017901 (2003)}.

\bibitem{TrappedIons_Schneider:2012}
C.~Schneider, D.~Porras, and T.~Schaetz, ``Experimental quantum simulations of
  many-body physics with trapped ions,''
  \href{http://dx.doi.org/10.1088/0034-4885/75/2/024401}{{\em Rep. Prog. Phys.}
  {\bfseries 75}, 024401 (2012)}.

\bibitem{safronova2018search}
M.~S. Safronova, D.~Budker, D.~DeMille, D.~F.~J. Kimball, A.~Derevianko, and
  C.~W. Clark, ``Search for new physics with atoms and molecules,''
  \href{http://dx.doi.org/10.1103/RevModPhys.90.025008}{{\em Rev. Mod. Phys.}
  {\bfseries 90}, 025008 (2018)}.

\bibitem{katori2011optical}
H.~Katori, ``Optical lattice clocks and quantum metrology,''
  \href{http://dx.doi.org/https://doi.org/10.1038/nphoton.2011.45}{{\em Nature
  Photonics} {\bfseries 5}, 203 (2011)}.

\bibitem{Hanneke2008}
D.~Hanneke, S.~Fogwell, and G.~Gabrielse, ``New Measurement of the Electron
  Magnetic Moment and the Fine Structure Constant,''
  \href{http://dx.doi.org/10.1103/PhysRevLett.100.120801}{{\em Phys. Rev.
  Lett.} {\bfseries 100}, 120801 (2008)}.

\bibitem{Odom2006}
B.~Odom, D.~Hanneke, B.~D'Urso, and G.~Gabrielse, ``New Measurement of the
  Electron Magnetic Moment Using a One-Electron Quantum Cyclotron,''
  \href{http://dx.doi.org/10.1103/PhysRevLett.97.030801}{{\em Phys. Rev. Lett.}
  {\bfseries 97}, 030801 (2006)}.

\bibitem{Gabrielse2006}
G.~Gabrielse, D.~Hanneke, T.~Kinoshita, M.~Nio, and B.~Odom, ``New
  Determination of the Fine Structure Constant from the Electron $g$ Value and
  QED,'' \href{http://dx.doi.org/10.1103/PhysRevLett.97.030802}{{\em Phys. Rev.
  Lett.} {\bfseries 97}, 030802 (2006)}.

\bibitem{Gogolin_2016}
C.~Gogolin and J.~Eisert, ``Equilibration, thermalisation, and the emergence of
  statistical mechanics in closed quantum systems,''
  \href{http://dx.doi.org/10.1088/0034-4885/79/5/056001}{{\em Reports on
  Progress in Physics} {\bfseries 79}, 056001 (2016)}.

\bibitem{Erne:420237}
S.~Erne, R.~B{\"u}cker, T.~Gasenzer, J.~Berges, and J.~Schmiedmayer,
  ``{U}niversal dynamics in an isolated one-dimensional {B}ose gas far from
  equilibrium,'' \href{http://dx.doi.org/10.1038/s41586-018-0667-0}{{\em
  Nature} {\bfseries 563}, 225 -- 229 (2018)}.

\bibitem{Einstein:1905}
A.~Einstein, ``Ist die Tr{\"a}gheit eines K{\"o}rpers von seinem Energieinhalt
  abh{\"a}ngig?,'' {\em Annalen der Physik} {\bfseries 323}, 639--641 (1905).

\bibitem{Einstein:1907}
A.~Einstein, ``{\"U}ber das Relativit{\"a}tsprinzip und die aus demselben
  gezogenen Folgerungen. Jahrb. f. Rad. und Elekt. 4, 411 (1907),'' {\em
  Jahrbuch der Radioaktivit{\"a}t} {\bfseries 4}, 411--462 (1907).

\bibitem{PhysRevLett.4.341}
B.~D. Josephson, ``Temperature-Dependent Shift of $\ensuremath{\gamma}$ Rays
  Emitted by a Solid,'' \href{http://dx.doi.org/10.1103/PhysRevLett.4.341}{{\em
  Phys. Rev. Lett.} {\bfseries 4}, 341--342 (1960)}.

\bibitem{Greenberger:1970_1}
D.~M. Greenberger, ``Theory of particles with variable mass. I. Formalism,''
  \href{http://dx.doi.org/10.1063/1.1665400}{{\em Journal of Mathematical
  Physics} {\bfseries 11}, 2329--2340 (1970)}.

\bibitem{Greenberger:1970_2}
D.~M. Greenberger, ``Theory of particles with variable mass. II. Some physical
  consequences,'' \href{http://dx.doi.org/10.1063/1.1665401}{{\em Journal of
  Mathematical Physics} {\bfseries 11}, 2341--2347 (1970)}.

\bibitem{Greenberger:1974}
D.~M. Greenberger, ``Some useful properties of a theory of variable mass
  particles,'' \href{http://dx.doi.org/10.1063/1.1666658}{{\em Journal of
  Mathematical Physics} {\bfseries 15}, 395--405 (1974)}.

\bibitem{Zych:2011}
M.~Zych, F.~Costa, I.~Pikovski, and {\v{C}}.~Brukner, ``Quantum interferometric
  visibility as a witness of general relativistic proper time,''
  \href{http://dx.doi.org/10.1038/ncomms1498}{{\em Nature Communications}
  {\bfseries 2}, 505 (2011)}.

\bibitem{Zych:2012}
M.~Zych, F.~Costa, I.~Pikovski, T.~C. Ralph, and {\v C}.~Brukner, ``General
  relativistic effects in quantum interference of photons,''
  \href{http://dx.doi.org/10.1088/0264-9381/29/22/224010}{{\em Classical and
  Quantum Gravity} {\bfseries 29}, 224010 (2012)}.

\bibitem{bushev2016single}
P.~A. Bushev, J.~H. Cole, D.~Sholokhov, N.~Kukharchyk, and M.~Zych, ``Single
  electron relativistic clock interferometer,''
  \href{http://dx.doi.org/10.1088/1367-2630/18/9/093050}{{\em New Journal of
  Physics} {\bfseries 18}, 093050 (2016)}.

\bibitem{pikovskiuniversal2015}
I.~Pikovski, M.~Zych, F.~Costa, and {\v C}.~Brukner, ``Universal decoherence
  due to gravitational time dilation,''
  \href{http://dx.doi.org/10.1038/nphys3366}{{\em Nature Physics} {\bfseries
  11}, 668--672 (2015)}.

\bibitem{PikovskiTime2017}
I.~Pikovski, M.~Zych, F.~Costa, and {\v C}.~Brukner, ``Time dilation in quantum
  systems and decoherence,''
  \href{http://dx.doi.org/10.1088/1367-2630/aa5d92}{{\em New Journal of
  Physics} {\bfseries 19}, 025011 (2017)}.

\bibitem{korbicz2017information}
J.~Korbicz and J.~Tuziemski, ``Information transfer during the universal
  gravitational decoherence,''
  \href{http://dx.doi.org/10.1007/s10714-017-2319-3}{{\em General Relativity
  and Gravitation} {\bfseries 49}, 152 (2017)}.

\bibitem{2016PhRvL.117i0401P}
B.~H. {Pang}, Y.~{Chen}, and F.~Y. {Khalili}, ``{Universal Decoherence under
  Gravity: A Perspective through the Equivalence Principle},''
  \href{http://dx.doi.org/10.1103/PhysRevLett.117.090401}{{\em Physical Review
  Letters} {\bfseries 117}, 090401 (2016)}.

\bibitem{Orlando2017}
P.~J. Orlando, F.~A. Pollock, and K.~Modi, {\em How Does Interference Fall?},
  \href{http://dx.doi.org/10.1007/978-3-319-53412-1_19}{pp.~421--451}.
\newblock Springer International Publishing, Cham, 2017.

\bibitem{paige2018quantum}
A.~Paige, A.~Plato, and M.~Kim, ``Quantum clocks do not witness classical time
  dilation,'' {\em arXiv preprint arXiv:1809.01517} (2018).

\bibitem{Sinha_2014}
S.~Sinha and J.~Samuel, ``Quantum limit on time measurement in a gravitational
  field,'' \href{http://dx.doi.org/10.1088/0264-9381/32/1/015018}{{\em
  Classical and Quantum Gravity} {\bfseries 32}, 015018 (2014)}.

\bibitem{Castro:2017clocks}
E.~Castro~Ruiz, F.~Giacomini, and {\v{C}}.~Brukner, ``Entanglement of quantum
  clocks through gravity,''
  \href{http://dx.doi.org/10.1073/pnas.1616427114}{{\em Proceedings of the
  National Academy of Sciences} {\bfseries 114}, E2303--E2309 (2017)}.

\bibitem{sonnleitner2017will}
M.~Sonnleitner, N.~Trautmann, and S.~M. Barnett, ``Will a decaying atom feel a
  friction force?,''
  \href{http://dx.doi.org/10.1103/PhysRevLett.118.053601}{{\em Physical Review
  Letters} {\bfseries 118}, 053601 (2017)}.

\bibitem{sonnleitner2018mass}
M.~Sonnleitner and S.~M. Barnett, ``Mass-energy and anomalous friction in
  quantum optics,'' \href{http://dx.doi.org/10.1103/PhysRevA.98.042106}{{\em
  Physical Review A} {\bfseries 98}, 042106 (2018)}.

\bibitem{ZychGreenberger2019}
M.~Zych and D.~M. Greenberger, ``Puzzling out the mass-superselection rule,''
  {\em arXiv preprint arXiv:1906.xxxx} (2019).

\bibitem{krause2016taking}
D.~E. Krause and I.~Lee, ``Taking Einstein seriously: Relativistic coupling of
  internal and center of mass dynamics,''
  \href{http://dx.doi.org/10.1088/1361-6404/aa6903}{{\em Eur.~J.~Phys.}
  {\bfseries 38}, 045401 (2017)}.

\bibitem{PhysRevLett.104.070802}
C.~W. Chou, D.~B. Hume, J.~C.~J. Koelemeij, D.~J. Wineland, and T.~Rosenband,
  ``Frequency Comparison of Two High-Accuracy ${\mathrm{Al}}^{+}$ Optical
  Clocks,'' \href{http://dx.doi.org/10.1103/PhysRevLett.104.070802}{{\em Phys.
  Rev. Lett.} {\bfseries 104}, 070802 (2010)}.

\bibitem{PhysRevLett.116.063001}
N.~Huntemann, C.~Sanner, B.~Lipphardt, C.~Tamm, and E.~Peik, ``Single-Ion
  Atomic Clock with
  $3\ifmmode\times\else\texttimes\fi{}{10}^{\ensuremath{-}18}$ Systematic
  Uncertainty,'' \href{http://dx.doi.org/10.1103/PhysRevLett.116.063001}{{\em
  Phys. Rev. Lett.} {\bfseries 116}, 063001 (2016)}.

\bibitem{PhysRevLett.118.053002}
J.-S. Chen, S.~M. Brewer, C.~W. Chou, D.~J. Wineland, D.~R. Leibrandt, and
  D.~B. Hume, ``Sympathetic Ground State Cooling and Time-Dilation Shifts in an
  ${^{27}\mathrm{Al}}^{+}$ Optical Clock,''
  \href{http://dx.doi.org/10.1103/PhysRevLett.118.053002}{{\em Phys. Rev.
  Lett.} {\bfseries 118}, 053002 (2017)}.

\bibitem{Yudin:2018}
V.~Yudin and A.~Taichenachev, ``Mass defect effects in atomic clocks,''
  \href{http://dx.doi.org/10.1088/1612-202x/aa9aa5}{{\em Laser Physics Letters}
  {\bfseries 15}, 035703 (2018)}.

\bibitem{zych2015PhD}
M.~Zych, {\em Quantum systems under gravitational time dilation}.
\newblock Springer Theses. Springer, 2017.

\bibitem{zych2015quantumEEP}
M.~Zych and {\v C}.~Brukner, ``Quantum formulation of the Einstein Equivalence
  Principle,'' \href{http://dx.doi.org/10.1038/s41567-018-0197-6}{{\em Nature
  Physics} {\bfseries 14}, 1027--1031 (2018)}.

\bibitem{Anastopoulos:2018}
C.~Anastopoulos and B.~L. Hu, ``Equivalence principle for quantum systems:
  dephasing and phase shift of free-falling particles,''
  \href{http://dx.doi.org/10.1088/1361-6382/aaa0e8}{{\em Classical and Quantum
  Gravity} {\bfseries 35}, 035011 (2018)}.

\bibitem{zych2018gravitational}
M.~Zych, L.~Rudnicki, and I.~Pikovski, ``Gravitational mass of composite
  systems,'' \href{http://dx.doi.org/10.1103/PhysRevD.99.104029}{{\em Phys.
  Rev. D} {\bfseries 99}, 104029 (2019)}.

\bibitem{Wineland:2010}
C.-W. Chou, D.~Hume, T.~Rosenband, and D.~Wineland, ``Optical clocks and
  relativity,'' \href{http://dx.doi.org/10.1126/science.1192720}{{\em Science}
  {\bfseries 329}, 1630--1633 (2010)}.

\bibitem{Reinhardt2007test}
S.~Reinhardt, G.~Saathoff, H.~Buhr, L.~A. Carlson, A.~Wolf, D.~Schwalm,
  S.~Karpuk, C.~Novotny, G.~Huber, M.~Zimmermann, {\em et al.}, ``Test of
  relativistic time dilation with fast optical atomic clocks at different
  velocities,'' \href{http://dx.doi.org/doi:10.1038/nphys778}{{\em Nature
  Physics} {\bfseries 3}, 861--864 (2007)}.

\bibitem{brewer201927}
S.~Brewer, J.-S. Chen, A.~Hankin, E.~Clements, C.~Chou, D.~Wineland, D.~Hume,
  and D.~Leibrandt, ``An $^{27}$ Al$^{+}$ quantum-logic clock with systematic
  uncertainty below $10^{-18}$,'' {\em arXiv preprint arXiv:1902.07694} (2019).

\bibitem{schleich1987oscillations}
W.~Schleich and J.~A. Wheeler, ``Oscillations in photon distribution of
  squeezed states and interference in phase space,''
  \href{http://dx.doi.org/10.1038/326574a0}{{\em Nature} {\bfseries 326}, 574
  (1987)}.

\bibitem{KimKnight:1989PRA}
M.~S. Kim, F.~A.~M. de~Oliveira, and P.~L. Knight, ``Properties of squeezed
  number states and squeezed thermal states,''
  \href{http://dx.doi.org/10.1103/PhysRevA.40.2494}{{\em Phys. Rev. A}
  {\bfseries 40}, 2494--2503 (1989)}.

\bibitem{toth2014quantum}
G.~T{\'o}th and I.~Apellaniz, ``Quantum metrology from a quantum information
  science perspective,''
  \href{http://dx.doi.org/10.1088/1751-8113/47/42/424006}{{\em Journal of
  Physics A: Mathematical and Theoretical} {\bfseries 47}, 424006 (2014)}.

\bibitem{Paris:2009Qest}
M.~Paris, ``Quantum estimation for quantum technology,''
  \href{http://dx.doi.org/10.1142/S0219749909004839}{{\em International Journal
  of Quantum Information} {\bfseries 07}, 125--137 (2009)}.

\bibitem{NolanHaine:2017PRA}
S.~P. Nolan and S.~A. Haine, ``Quantum Fisher information as a predictor of
  decoherence in the preparation of spin-cat states for quantum metrology,''
  \href{http://dx.doi.org/10.1103/PhysRevA.95.043642}{{\em Phys. Rev. A}
  {\bfseries 95}, 043642 (2017)}.

\bibitem{Milburn:1994Squeeze}
G.~J. Milburn, W.-Y. Chen, and K.~R. Jones, ``Hyperbolic phase and
  squeeze-parameter estimation,''
  \href{http://dx.doi.org/10.1103/PhysRevA.50.801}{{\em Phys. Rev. A}
  {\bfseries 50}, 801--804 (1994)}.

\bibitem{Chiribella:2006Squeezing}
G.~Chiribella, G.~M. D'Ariano, and M.~F. Sacchi, ``Optimal estimation of
  squeezing,'' \href{http://dx.doi.org/10.1103/PhysRevA.73.062103}{{\em Phys.
  Rev. A} {\bfseries 73}, 062103 (2006)}.

\bibitem{Gaertner:2017OTOC}
M.~G{\"a}rttner, J.~G. Bohnet, A.~Safavi-Naini, M.~L. Wall, J.~J. Bollinger,
  and A.~M. Rey, ``Measuring out-of-time-order correlations and multiple
  quantum spectra in a trapped-ion quantum magnet,''
  \href{http://dx.doi.org/10.1038/nphys4119}{{\em Nature Physics} {\bfseries
  13}, 781--786 (2017)}.

\bibitem{LewisSwan:2019FOTOC}
R.~J.~Lewis-Swan, A.~Safavi-Naini, J.~Bollinger, and A.~M.~Rey, ``Unifying
  scrambling, thermalization and entanglement through measurement of fidelity
  out-of-time-order correlators in the Dicke model,''
  \href{http://dx.doi.org/10.1038/s41467-019-09436-y}{{\em Nature
  Communications} {\bfseries 10}, 1581 (2019)}.

\bibitem{roura2018gravitational}
A.~Roura, ``Gravitational redshift in quantum-clock interferometry,'' {\em
  arXiv preprint arXiv:1810.06744} (2018).

\bibitem{2016CQGra..33sLT01O}
P.~J. {Orlando}, R.~B. {Mann}, K.~{Modi}, and F.~A. {Pollock}, ``{A test of the
  equivalence principle(s) for quantum superpositions},''
  \href{http://dx.doi.org/10.1088/0264-9381/33/19/19LT01}{{\em Classical and
  Quantum Gravity} {\bfseries 33}, 19LT01 (2016)}.

\bibitem{geiger2018proposal}
R.~Geiger and M.~Trupke, ``Proposal for a Quantum Test of the Weak Equivalence
  Principle with Entangled Atomic Species,''
  \href{http://dx.doi.org/10.1103/PhysRevLett.120.043602}{{\em Physical Review
  Letters} {\bfseries 120}, 043602 (2018)}.

\bibitem{rosi2017quantum}
G.~Rosi, G.~D'Amico, L.~Cacciapuoti, F.~Sorrentino, M.~Prevedelli, M.~Zych,
  {\v{C}}.~Brukner, and G.~Tino, ``Quantum test of the equivalence principle
  for atoms in coherent superposition of internal energy states,''
  \href{http://dx.doi.org/10.1038/ncomms15529}{{\em Nature Communications}
  {\bfseries 8}, 15529 (2017)}.

\bibitem{Anders_2017}
J.~Anders and M.~Esposito, ``Focus on quantum thermodynamics,''
  \href{http://dx.doi.org/10.1088/1367-2630/19/1/010201}{{\em New Journal of
  Physics} {\bfseries 19}, 010201 (2017)}.

\bibitem{binder2018thermodynamics}
F.~Binder, L.~A. Correa, C.~Gogolin, J.~Anders, and G.~Adesso,
  \href{http://dx.doi.org/10.1007/978-3-319-99046-0}{{\em Thermodynamics in the
  quantum regime}}.
\newblock Thermodynamics in the Quantum Regime: Fundamental Aspects and New
  Directions. Springer, 2018.

\bibitem{PhysRevD.23.2157}
E.~Fischbach, B.~S. Freeman, and W.-K. Cheng, ``General-relativistic effects in
  hydrogenic systems,'' \href{http://dx.doi.org/10.1103/PhysRevD.23.2157}{{\em
  Physical Review D} {\bfseries 23}, 2157--2180 (1981)}.

\bibitem{Nordtvedt:1994PostNewtonian}
K.~Nordtvedt, ``Post-Newtonian gravity: its theory--experiment interface,''
  \href{http://dx.doi.org/https://doi.org/10.1088/0264-9381/11/6A/009}{{\em
  Classical and Quantum Gravity} {\bfseries 11}, A119 (1994)}.

\bibitem{Carlip:1999KE}
S.~Carlip, ``Kinetic energy and the equivalence principle,''
  \href{http://dx.doi.org/10.1119/1.18885}{{\em American Journal of Physics}
  {\bfseries 66}, 409--413 (1998)}.

\bibitem{Gersch:1992minimal}
H.~A. Gersch, ``Time evolution of minimum uncertainty states of a harmonic
  oscillator,'' \href{http://dx.doi.org/10.1119/1.16981}{{\em American Journal
  of Physics} {\bfseries 60}, 1024--1030 (1992)}.

\end{thebibliography}\endgroup

\section{Supplementary Material for ``Mass-energy equivalence in harmonically trapped particles''}
\subsection{Derivation of the Hamiltonian}
We are interested in the Hamiltonian describing a bound system in a weak gravitational field. Denoting the local, rest frame  Hamiltonian by $H_{loc}$ the coupling to gravity up to first post-Newtonian order gives the total Hamiltonian $H=H_{loc}(1+\frac{\phi(x)}{c^2})$ where $\phi(x)$  is the gravitational potential and $x$ is the external coordinate~\cite{zych2018gravitational}. We  also assume that the local frame is stationary in the gravitational field, which is satisfied for a laboratory experiment on Earth.
In the scenario considered, we have a low-energy composite particle in an external potential. Due to the assumption of the weak gravitational field the low-energy condition holds in the local rest frame as well as in the laboratory reference frame. The local Hamiltonian thus takes the form $H_{loc}\approx Mc^2+\frac{p_0^2}{2M}+V_0(x_0)$ where $V_{0}(x_0)$ is the potential in the local rest frame and $x_0$ is the corresponding coordinate, which is related to the laboratory coordinate $x$ as $x_0=x(1-\frac{\phi(x)}{c^2})$. Similarly, $p_0$ is the local momentum related to the momentum $p$ in the laboratory frame as $p_0=p(1+\frac{\phi(x)}{c^2})$. Since we need the Hamiltonian in the laboratory frame we transform also $H_{loc}$ to the laboratory coordinates. It is meaningful to consider the potential $V_0(x_0)$ as coming from $1/x_0$ interactions, in which case it transforms as $V_0(x_0)=V(x)(1+\frac{\phi(x)}{c^2})$. (As a toy model consider Coulomb forces between two fixed charges at a local distance $2d_0$ which produce an approximately harmonic potential for a test charge between them: $V_0(x_0)\propto\frac{1}{d_0+x_0}+\frac{1}{d_0-x_0}\approx  \mathrm{const}+2\frac{x_0^2}{d_0^3}$.)
 A simple substitution gives 
 \be
 \begin{split}
 H=& Mc^2(1+\frac{\phi(x)}{c^2})\\+&\frac{p^2}{2M}(1+3\frac{\phi(x)}{c^2})+V(x)(1+2\frac{\phi(x)}{c^2}),
 \end{split}
 \ee{fullHam} 
 which has a general form as expected in this regime from other studies \cite{PhysRevD.23.2157, Nordtvedt:1994PostNewtonian,Carlip:1999KE}.
The correction $p^2\phi(x)/2Mc^2$ is of higher order than the desired Newtonian limit for the CM, and is thus neglected. Furthermore, taking the external potential to be approximately harmonic in the local frame $V_0(x_0)\approx \frac{k_0}{2}x_0^2$, we obtain a slightly anharmonic potential in the laboratory frame -- this anharmonicity is, however, suppressed by $1/c^2$ and will thus be much smaller than real anharmonicities of laboratory trapping potentials. With these in mind, the laboratory frame Hamiltonian in \eq{fullHam} reduces to \eq{Ham} in the main text for $V(x)=kx^2/2$. 

Let us stress that the trap stiffness $k$ in \eq{Ham} is therefore the laboratory frame stiffness. For a trap at a different height $x+h$ in the laboratory frame, we have $(1+2\frac{\phi(x+h)}{c^2})\approx (1+2\frac{\phi(x)}{c^2}+2\frac{g h}{c^2})\approx(1+2\frac{\phi(x)}{c^2})(1+2\frac{gh}{c^2})$ where $g$ is the gravitational acceleration at $x$.  We thus obtain that for two \textit{locally identical} traps located at heights differing by $h$ the laboratory observer will  assign stiffness parameters that differ by a factor $(1+2\frac{gh}{c^2})$. As a result, the CM frequencies $\sqrt{k/M}$ seen in the laboratory frame will differ by the gravitational redshift factor $(1+\frac{gh}{c^2})$ as expected.

Finally, note that \eq{Ham} can be formally diagonalised as follows
\be
H=\h Mc^2(1-\frac{g^2}{2\ohat^2c^2})+\hbar\ohat\left( \h a_{\h M}^\dagger\h a_{\h M}+\frac{1}{2}\right),
\ee{HaM}
where $\h a_{\h M}  := \sum_i\h a_{i}\h\Pi_i$,  $\ohat:=({k/\h M})^{1/2}$, and ${1/\h M}=\sum_i({1/M_i})\h\Pi_i$ (which is well-defined as $\forall_i M_i>0$). In this notation $\h a_{\h M} = ({{\Mhat\ohat}/{2\hbar}})^{1/2}(\xhat +g/\ohat^2+ {i}\phat/{\Mhat \ohat})$.

\subsection{Ramsey spectroscopy}
From the canonical transformation in \eq{squeezing} immediately follows that the Fock bases  associated with different internal states are related as $\ket{n_0}=D(\alpha_{gi})S(r_i)\ket{n_i}$. For $\h \rho_0=\ket{\psi_0}\bra{\psi_0}$, the trace in \eq{probab} can thus be expressed as  $\bra{\psi_0}\h U_0(t)^\dagger \h U_1(\omega_1,t)D(\alpha_{g})S(r)\ket{\psi_1}$ and we used $\ket{\psi_0}=D(\alpha_{g})S(r)\ket{\psi_1}$. For $\ket{\psi_0}$ the CM ground state associated with internal energy $E_0$ this further simplifies to $\bra{\psi_0} \h U_1(\omega_1,t)D(\alpha_{g})S(r)\ket{\psi_1}$ and we thus need to evaluate an amplitude between $\ket{\psi_0}$ and time evolved displaced squeezed vacuum state of an oscillator with mass $M_1$, \ie $\ket{\psi_1(t)}=\h U_1(\omega_1,t)D(\alpha_{g})S(r)\ket{\psi_1} $. Following e.g.~ref.~\cite{Gersch:1992minimal}  we find  
\be
\psi_1(x,t) = \sqrt[4]{\frac{a_0}{\pi}}\frac{e^{-i\phi_1(t)-\frac{a_0(R(t)+i I(t))}{2 (S^2 + (1 - S^2) \cos^2(\omega_1 t))}}}{\sqrt{\cos( \omega_1t)+i S \sin(\omega_1t)}}
\ee{dsdvac}
where $\phi_i(t) := M_ic^2(1-\frac{g^2}{2\omega_i^2c^2})t,\;i=0,1$,  $R(t):=(x - x_0 \cos(\omega_1 t))^2$, $I(t):=\sin(\omega_1 t) \{S (x^2 + x_0^2)\cos(\omega_1 t) - 2 S x x_0 - \frac{x^2 \cos(\omega_1 t)}{S}\}$, 
with $S=\sqrt{\frac{M_0}{M_1}}$, $x_0=-\frac{g}{\omega_0^2}$ and $a_0=\frac{M_0\omega_0}{\hbar}$. 
 The full trace is $\int dx\psi^*_0(x,t)\psi_1(x,t)\equiv\mathcal{V}_{\mathrm{vac}}(t)e^{i\varphi(t)}$ where $\psi_0(x,t)= \sqrt[4]{\frac{a_0}{\pi}}e^{-i\phi_0(t)-i\frac{\omega_0}{2}t-\frac{a(x-x_0)^2}{2}}$,
\be
\mathcal{V}_{\mathrm{vac}}(t) = \frac{\sqrt{2S}\,e^{\frac{- a x_0^2}{1 + S^2 \cot^2(\omega_1t/2)}}}{\sqrt[4]{4 S^2 + (1 - S^2)^2 \sin^2(\omega_1t)}} 
\ee{visibvac}
and
\be
\varphi(t) =  {\omega_0t}/{2}+\Delta\phi+\varphi/2+\frac{a x_0^2 \,S \cot(\omega_1t/2)}{1 + S^2\cot^2(\omega_1t/2)},
\ee{phase_vac}
with  $\Delta\phi :=\phi_0(t)-\phi_1(t)$ and $\varphi:= \mathrm{Arg}[i(S^2-1) \sin(t \omega_1) + \frac{2 S}{\cos(t \omega_1) -i S \sin(t \omega_1)}]$.
From \eq{visibvac} we find that the (first) partial revival occurs at $t_{rev}= \frac{\pi}{\omega_1}$ where $\mathcal{V}_{\mathrm{vac}}(t_{rev})=  e^{-a x_0^2}$. The analytical expression for the first minimum can also be obtained, but due to its complicated form we state here only the result for small $x_0$ and $S$, where the minimum is reached for $t_{min}\approx \frac{\pi}{2\omega_1}$ and reads $\mathcal{V}_{\mathrm{vac}}(t_{min})=  e^{\frac{-a x_0^2}{1 + S^2}} \sqrt{\frac{2S}{1 + S^2}}$.

\begin{figure}[h!]
\includegraphics[width=0.9\columnwidth]{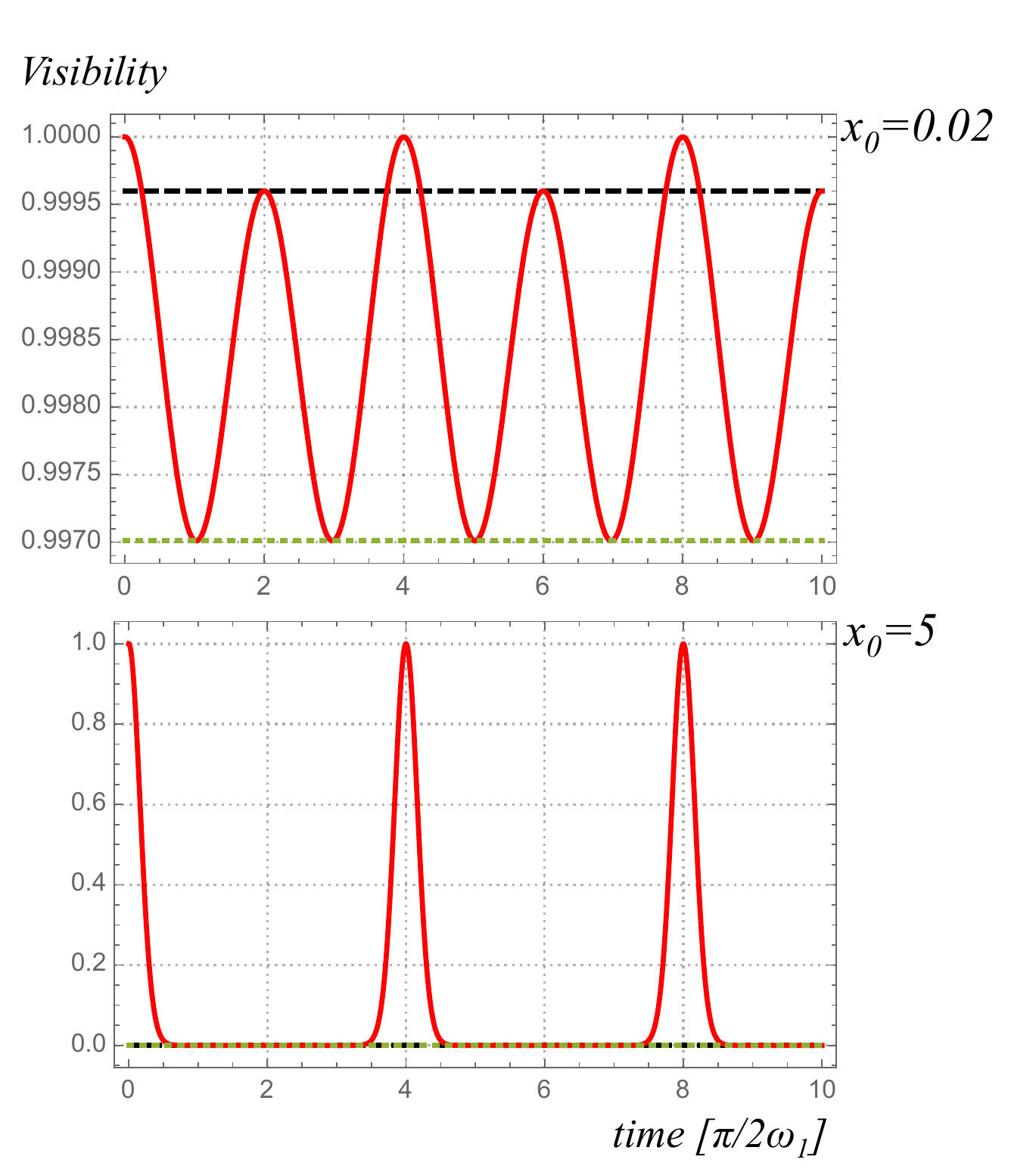}
\caption{Interference visibility for the coherent states of different magnitude top: $x_0=0.02$, bottom: $x_0=5$. Thick red line is the full visibility \eq{visibvac}, dotted green line is the minimum visibility $\mathcal{V}(t_\mathrm{min})= e^{\frac{-a_0 x_0^2}{1 + S^2}} \sqrt{\frac{2S}{1 + S^2}}$ and dashed black line is the visibility at the revival time $\mathcal{V}(t_{\mathrm{rev}})= e^{-a_0 x_0^2}$. Other parameters are $S:=\sqrt{M_0/M_1} = 0.9$, and $ a_0:=M_0\omega_0/\hbar=1$.}
\label{SuppFig1}
\end{figure}

The above results for an arbitrary $x_0$ (as opposed to an exact value $x_0=-g/\omega_0^2$) give the interference for an initial coherent state rather than the vacuum. The general features of the interference pattern are analogous: the times $t_{min}, t_{rev}$ are the same as for the vacuum but for larger $|x_0|$ the visibility has a faster drop, the revivals are shorter, and the  values of the visibility at the minima and at the partial revivals are smaller, see \fig{SuppFig1}.  Finally, note that neglecting gravity gives $x_0=0$ and the visibility has a complete revival at $t=t_{rev}$. 

\subsection{Key steps in derviations}
Here we give the key formulas to reproduce the results from the main text. We mostly make use of matrix elements or expectation values of number operator $ \h n_k $ which reads
\be
 \begin{split}
 \h n_k = &\h n_0+\sinh^2(r_k) - \frac{\sinh(2r_k)}{2}(\h a_0^2 +\h a_0^{\dagger2})\\ 
 +&\alpha_{gk}e^{-r_k}(\h a_0^\dagger+\h a_0)+\alpha_{gk}^2.
\end{split}
\ee{number}

To obtain the explicit form of the phase shift, the formulas just after \eq{full_shift}, we need $\mean{\h n_1}-\mean{\h n_0} = \sinh^2(r)-\frac{1}{2}\sinh(2r)\left(\langle \h a^{\dagger2}_0\rangle +\mean{a_0^{2}}\right) $ and $\langle[\h n_0, \h n_1]\rangle=\sinh(2r)(\mean{\h a_0^2}-\langle \h a^{\dagger2}_0\rangle)+\alpha_g e^{-r}(\h a_0^\dagger-\h a_0)$.  When inserted in \eq{full_shift} these expressions directly  give the parameters $A_0, A_g$ from the main text.

In order to reproduce the results for the time evolution of a coherent state (\eq{Qfunct} and  discussion below) the key expression is $\frac{\bra{\beta}\h n_k^2\ket{\alpha}}{\braket{\beta| \alpha}}-\big(\frac{\bra{\beta}\h n_k\ket{\alpha}}{\braket{\beta|\alpha}}\big)^2= {\beta^2}\frac{E_k}{2M_0c^2} + \beta \alpha_{gk} +\mathcal{O}((\frac{E_k}{M_0c^2})^2)$.   \eq{alpha_g} yields $\alpha_{gk}\approx \frac{g E_k}{c^2\sqrt{2\hbar M_0\omega_0^2}}$; expanding the exponential in \eq{Qfunct} and keeping its first two terms further gives $\sum_kp_ke^{-(\omega_kt)^2\big(\frac{\bra{\beta}\h n_k^2\ket{\alpha}}{\braket{\beta| \alpha}}-\big(\frac{\bra{\beta}\h n_k\ket{\alpha}}{\braket{\beta|\alpha}}\big)^2\big)}\approx$ $$1-(\omega_0t)^2\sum_k p_k  \frac{E_k}{2M_0c^2} \left({\beta^2} + \beta g\sqrt{\frac{2 M_0}{\hbar\omega_0^3}}\right).$$ Using that $\sum_kp_kE_k\equiv\langle\h H_{int}\rangle$ gives the expression provided in the main text.

\end{document}